\documentclass[%
 twocolumn,
 reprint,
showpacs,
 amsmath,amssymb,
 aps,
]{revtex4}

\usepackage{color}
\usepackage{ulem}
\usepackage{graphicx}% Include figure files
\usepackage{dcolumn}% Align table columns on decimal point\usepackage{bm}% bold math
\usepackage{bm}% bold math
\usepackage{here}

\begin{document}

\preprint{APS/123-QED\author{Takahiro Ohgoe}}

\title{Variational Monte Carlo Method for Electron-Phonon Coupled Systems}% Force line breaks with \\

\author{Takahiro Ohgoe}
\author{Masatoshi Imada}%
\affiliation{%
	Department of Applied Physics, University of Tokyo, 7-3-1 Hongo, Bunkyo-ku, Tokyo 113-0033, Japan
}%

\date{\today}% It is always \today, today,

\begin{abstract}
  We develop a variational Monte Carlo (VMC) method for electron-phonon coupled systems. The VMC method has been extensively used for investigating strongly correlated electrons over the last decades. However, its applications to electron-phonon coupled systems have been severely restricted because of its large Hilbert space. Here, we propose a variational wave function with a large number of variational parameters which is suitable and tractable for systems with electron-phonon coupling. In the proposed wave function, we implement an unexplored electron-phonon correlation factor which takes into account the effect of the entanglement between electrons and phonons. The method is applied to systems with diagonal electron-phonon interactions, i.e. interactions between charge densities and lattice displacements (phonons). As benchmarks, we compare VMC results with previous results obtained by the exact diagonalization, the Green function Monte Carlo and the density matrix renormalization group for the Holstein and Holstein-Hubbard model. From these benchmarks, we show that the present method offers an efficient way to treat strongly coupled electron-phonon systems.

%PACS numbers: May be entered using the \verb+\pacs{#1}+ command.
\end{abstract}

\maketitle

%\tableofcontents

\section{\label{sec:intro}Introduction}
The electron-phonon coupling plays an important role in various classes of materials. In the conventional superconductors, it is the origin of the effective attraction between electrons, which leads to the formation of Cooper pairs\cite{bardeen1957}. Even for high-$T_c$ cuprates, the ARPES (angle-resolved photoemission spectroscopy) experiments demonstrated indications of strong electron-phonon coupling\cite{lanzara2001}. Apart from the superconductors, in quasi-one-dimensional materials, the electron-phonon coupling sometimes drives the Peierls transition\cite{peierls1955}, where the lattice is deformed and the electrons become insulating with charge density wave (CDW).

There are several numerical methods to tackle the problems of electron-phonon coupled systems such as the exact diagonalization (ED)\cite{marsiglio1993,marsiglio1995}, the density matrix renormalization group (DMRG)\cite{white1992,jeckelmann1998,bursill1998,tezuka2005,tezuka2007,fehske2008,ejima2009}, the quantum Monte Carlo (QMC) method\cite{hirsch1983,scalettar1989,noack1991,niyaz1993,mishchenko2000,hohenadler2004,spencer2005,assaad2008,hohenadler2012,nowadnick2012,johnston2013}, the dynamical mean-field theory (DMFT)\cite{georges1996,sangiovanni2005,sangiovanni2006,p_werner2007,p_werner2013,murakami2013}, and so on. Although the ED provides exact results, it is limited to finite clusters. The DMRG is the most successful method to investigate the ground-state properties of one-dimensional systems with short-range interactions. By using the DMRG, ground-state phase diagrams of the Holstein-Hubbard model in one dimension have been obtained\cite{tezuka2007,fehske2008}. In contrast, the DMFT becomes exact in inifinite dimensions\cite{georges1996}. Since the DMFT neglects the spatial correlation, the DMFT studies have been devoted to the model with on-site (Holstein-type) electron-phonon interactions\cite{holstein1959}. The QMC method provides numerically exact results and various QMC methods have been developed for models such as the Su-Schrieffer-Heeger (SSH) model\cite{su1979} and the Fr${\ddot{\rm o}}$hlich model\cite{frohlich1950} as well as the Holstein model. However, the applications of the QMC methods are restricted to some parameter regions (dilute limit, at half-filling, or small system sizes) due to the notorious negative sign problem. In addition to these numerical methods, some variational approachs have been successful in one dimensions\cite{wellein1998,bonca1999,barisic2002,cataudella2004,barone2007,hardy2009,chakraborty2013,chakraborty2014} or infinite dimensions\cite{barone2008}. For polaron systems, even two and three dimensions are accessible\cite{ku2002,berciu2007,li2010}.

 Since the variational Monte Carlo (VMC) method does not suffer from the negative sign problem, it has been extensively used for investigating strongly correlated electrons over the last decades. However, most of its applications are restricted to systems without the electron-phonon coupling or effective electron models in the antiadiabatic limit\cite{hardy2009}. The main reason for this may be the difficulty of constructing a suitable and tractable variational wave functions with a small number of variational parameters for such systems. To our knowledge, the only attempt was made by Alder {\it et al.}\cite{alder1997} who specifically treated the model with the off-diagonal electron-phonon interaction, namely the hopping amplitude which depends on the lattice displacement\cite{su1979}. However, their variational wave function does not take into account the excited states of phonons which are not negligible for systems with strong electron-phonon interactions. Furthermore, the one-body part of their electron wave function is similar to a Fermi sea which lacks the accuracy and flexibility enough to describe different phases.

Although the VMC method allows us to perform simulations of large systems of electrons, the main drawback is the presence of the bias which the assumed variational wave function inherently has. However, in recent years, it has become possible to reduce the bias by largely increasing the number of variational parameters in the wave functions and by optimizing them simultaneously\cite{sorella2001,casula2004,sorella2005,sorella2007,tahara2008}.

In this paper, we show that this development also opens a way of applying the VMC method to electron-phonon coupled systems by proposing a suitable and tractable variational wave function with a large number of variational parameters. To treat the effect of the entanglement between electrons and phonons in an efficient way, we include an unexplored electron-phonon correlation factor in the proposed variational wave function. The benchmark results compared with numerically exact results demonstrate the accuracy of the method. Since the present method can be flexibly applied to large systems with any lattice structure and spatial dimensionality, the method will offer a way of treating so far difficult problems in systems with strong electron-phonon and electron-electron interactions.

The paper is organized as follows. In Sec. \ref{sec:wf}, we introduce a variational wave function with a large number of variational parameters for electron-phonon coupled systems. Sec. \ref{sec:vmc} describes how to implement the VMC based on our variational wave function. In Sec. \ref{sec:res}, we present benchmark results to show the accuracy of our variational wave function. Finally, we summarize our results in Sec. \ref{sec:sum}.

\section{\label{sec:wf}Variational Wave Function for Electron-Phonon Coupled Systems}
In this section, we propose a variational wave function for electron-phonon coupled systems. The general wave function can be constructed as a superposition of the tensor product states of all electron Fock states and all phonon Fock states. In general, such a wave function is not tractable because the number of basis grows exponentially as the system size increases. Instead, we take the way of constructing a varitational wave function as a tensor product state of an electron wave function and a phonon wave function with variational parameters. However, such a wave function does not take into account the effect of the entanglement due to the electron-phonon interactions. To include this effect, we introduce a correlation factor between electrons and phonons. Namely, our variatonal wave function takes the following form:
\begin{eqnarray}
  | \psi \rangle = {\cal P}^{\rm e-ph} (| \psi^{\rm ph} \rangle | \psi^{\rm ele} \rangle ),
\end{eqnarray}
where ${\cal P}^{\rm e-ph}$ is the electron-phonon correlation factor, $| \psi^{\rm ele} \rangle$ is an electron wave function and $| \psi^{\rm ph} \rangle$ is a phonon wave function. In the subsections below, we present the explicit form of $| \psi^{\rm ele} \rangle$, $| \psi^{\rm ph} \rangle$, and ${\cal P}^{\rm e-ph}$, which are suitable for the Hubbard model with the diagonal electron-phonon interactions, i.e. the interaction between charge densities and lattice displacements. The Holstein-Hubbard model\cite{holstein1959} and the Fr$\ddot{{\rm o}}$hlich-Hubbard model\cite{frohlich1950} belong to this class.

\subsection{\label{sec:wf_ele}Variational wave functions for electron part}
For an electron variational wave function, we adopt the following form:
\begin{eqnarray}
  | \psi^{\rm ele} \rangle = {\cal P}^{\rm J} {\cal P}^{\rm G} | \phi^{\rm pair} \rangle,
\end{eqnarray}
where the one-body part is the singlet paring wave function\cite{giamarchi1991,himeda2000,tahara2008} given by
\begin{eqnarray}
  | \phi^{\rm pair} \rangle = \left( \sum_{i,j=1}^{N} f_{ij} c_{i \uparrow}^{\dagger} c_{j \downarrow}^{\dagger}  \right)^{N_{\rm e}/2} | 0 \rangle, \label{eq:paring_wf}
\end{eqnarray}
and the projection operators are the Gutzwiller factor\cite{gutzwiller1963} and the Jastrow factor\cite{jastrow1955} given by
\begin{eqnarray}
  {\cal P}^{\rm G} & = & \exp \left( \sum_{i} \alpha_i^{\rm G} n_{i \uparrow} n_{i \downarrow} \right),\\
  {\cal P}^{\rm J} & = & \exp \left( \sum_{i<j} \alpha_{ij}^{\rm J} n_{i} n_{j} \right),
\end{eqnarray}
respectively. Here, $N$ and $N_{\rm e}$ are the number of sites and electrons, respectively. $c_{i \sigma}(c_{i \sigma}^{\dagger})$ represents the annihilation (creation) operator of an electron with spin $\sigma$ (=$\uparrow$ or $\downarrow$) at a site $i$. The particle number operators $n_{i \sigma}$ and $n_{i}$ are defined by $n_{i \sigma}=c_{i \sigma}^{\dagger} c_{i \sigma}$ and $n_i=n_{i \uparrow}+n_{i \downarrow}$. The variational parameters are $f_{ij}$, $\alpha_i^{\rm G}$ and $\alpha_i^{\rm J}$. The number of variational parameters for $| \phi^{\rm pair} \rangle$, ${\cal P}^{\rm G}$, and ${\cal P}^{\rm J}$ are $O(N^2)$, $O(N)$, and $O(N^2)$, respectively. However, one can reduce to $O(N)$, $O(1)$, and $O(N)$, respectively if we assume a sub-lattice structure.

The paring wave function $| \phi^{\rm pair} \rangle$ has an extended form of the Hartree-Fock-Bogoliubov-type one with the antiferromagnetic and superconducting orders which was introduced in Ref. \cite{giamarchi1991}. Therefore, it can flexibly describe paramagnetic metals, the antiferromagnetic states, and the superconducting states. The Gutzwiller factor and the Jastrow factor take into account the correlation effects and thus we can include many-body effects beyond the mean-field level\cite{capello2005}. In the limit of $\alpha_i^{\rm G} \to -\infty$, the paring wave function with the Guztwiller factor can describe the resonating valence bond (RVB) wave function\cite{liang1988}. If necessary, we can extend the variational wave function by adding the doublon-holon correlation factor\cite{kaplan1982,yokoyama1990} and by introducing the quantum-number projection\cite{tahara2008}.

\subsection{\label{sec:wf_ph}Variational wave functions for phonon part}

We consider the Hamiltonian for phonons which is given by
\begin{eqnarray}
  {\cal H}_{\rm ph} & = & \frac{1}{2M} \sum_{\bm q} (\Pi_{\bm q} \Pi_{-{\bm q}} + M^2 \omega_{\bm q}^2 Q_{\bm q} Q_{-{\bm q}}) \\
  & = & \sum_{\bm q} \omega_{\bm q}  \left( b_{\bm q}^{\dagger} b_{\bm q} +\frac{1}{2} \right),
\end{eqnarray}
where $M$ is the mass of atoms, ${\bm q}$ is the wave number vector, $\omega_{\bm q}$ is the phonon frequency, $Q_{\bm q}$ is the normal coordinate operator of the lattice displacements, $\Pi_{\bm q}$ is its conjugate momentum operator and $b_{\bm q} (b_{\bm q}^{\dagger})$ is the annihilation(creation) operator. Throughout this paper, we set $\hbar$ to 1. When there is no electron-phonon coupling, the ground state is simply written as $\prod_{\bm q} |m_{\bm q}=0\rangle$, where $|m_{\bm q} \rangle$ represents the Fock state of phonons. However, in the presence of a electron-phonon couping, phonons occupy the excited states as well as the lowest energy states. In order to describe such excitations of phonons, we adopt the following form as the phonon variational wave function:
\begin{eqnarray}
  | \psi^{\rm ph} \rangle & = & \prod_{\bm q} | \psi_{\bm q}^{\rm ph} \rangle \\
  & = & \prod_{\bm q} \left( \sum_{m_{\bm q}=0}^{m_{\bm q}^{\rm max}} c_{m_{\bm q}} | m_{\bm q} \rangle \right),
  \label{phonon_wf}
\end{eqnarray}
where $\{ c_{m_{\bm q}} \}$ are coefficients of a superposition of the Fock states with a wave vector $q$ and $m_{q}^{\rm max}$ is the cutoff in $m_{\bm q}$. We treat the coefficients $\{ c_{m_{\bm q}} \}$ as variational parameters. The number of the variational parameters is $O(N m^{\rm max})$ if $m_{\bm q}^{\rm max} = m^{\rm max}$. In practice, we can take different values of the cutoffs depending on the wave vectors ${\bm q}$, because a large value of the cutoff is sometimes required for a particular wave vector ${\bm q}$. Here, Eq.(\ref{phonon_wf}) is formulated for systems with a single branch of phonons, while the possible extension to systems with several different branches of phonons will be discussed later.

As described later in Sec. \ref{sec:vmc}, we use the eigenstates of the normal coordinate operators as the basis of phonon Hilbert space in the VMC. However, the normal coordinate operators are non-Hermitian and thus a set of the eigenstates does not form a complete basis. Because of this, we use modified normal coordinates. Correspondingly, we modify our phonon variational wave function. These details are described in Appendix.

\subsection{\label{sec:wf_ep}Electron-phonon correlation factor}
We consider the diagonal electron-phonon interaction of the following form 
\begin{eqnarray}
  {\cal H}_{\rm e-ph} & = & \sum_{i,j} g_{ij} x_i n_j,
\end{eqnarray}
where $g_{ij}$ is the strength of the electron-phonon (lattice) interaction and $x_i$ is the lattice displacement at site $i$. This interaction can also be written in terms of the normal coordinates $\{ Q_{\bm q} \}$ through $x_i = \frac{1}{\sqrt{N}} \sum_{\bm q} Q_{\bm q} e^{i {\bm q} \cdot {\bm r}_i}$. Especially for systems with a translational symmetry, they can be simply written as  
\begin{eqnarray}
  {\cal H}_{\rm e-ph}  & = & \sqrt{N} \sum_{\bm q} g_{\bm q} n_{{- {\bm q}}} Q_{\bm q} \label{eq:Hep}
\end{eqnarray}
where $g_q$ and $n_q$ are the Fourier transformations of $g_{ij}$ and $n_i$, respectively. Eq. (\ref{eq:Hep}) is equivalent to more familiar form of $\sum_{{\bm q},{\bm k}} g_{\bm q} \sqrt{\frac{1}{2M\omega_{\bm q}}} c_{{\bm k}+{\bm q}}^{\dagger} c_{\bm k} (b_{\bm q} + b_{-{\bm q}}^{\dagger})$.

In order to include the effect of the entanglement between electrons and phonons, we introduce the following correlation factor (projection operator)
\begin{eqnarray}
  {\cal P^{\rm e-ph}} & = & \exp \left( \sum_{i,j} {\alpha}_{ij} x_i n_j \right) ,
\end{eqnarray}
where the coefficients $\{ \alpha_{ij} \}$ are variational parameters. The number of the variational parameters is $O(N^2)$, but one can reduce to $O(N)$ if we assume a sub-lattice structure. For systems with strong local electron-phonon interactions, the electron-displacement correlation function decays exponentially\cite{jeckelmann1998,wellein1998,bonca1999,ku2002}. In such a case, it is expected that long-range part of $\alpha_{ij}$ is negligible and the number of variational parameters reduce from $O(N)$ to $O(1)$. This correlation factor is similar to the conventional correlation factors such as the Gutzwiller factor in the sense that it can impose smaller weights on configurations with higher interaction energy. Therefore, we expect that we can reasonably include the correlation effect arising from the electron-phonon coupling.

The extension to systems with several different branches of phonon dispersions is straightforward. Namely, we can extend the phonon wave functions and the electron-phonon correlation factor to 
\begin{eqnarray}
  | \psi^{\rm ph} \rangle = \prod_{ {\bm q} \lambda} \left( \sum_{m_{ {\bm q} \lambda}=0}^{m_{{\bm q} \lambda}^{\rm max}} c_{m_{ {\bm q} \lambda}} | m_{ {\bm q} \lambda} \rangle \right),
\end{eqnarray}
and 
\begin{eqnarray}
  {\cal P^{\rm e-ph}} & = & \exp \left( \sum_{i,j,\lambda} {\alpha}_{ij \lambda} x_{i \lambda} n_j \right) ,
\end{eqnarray}
respectively. Here, $\lambda$ represents of a branch of the phonon dispersion. The number of the variational parameters increases linearly as the number of the branches increases.

\section{\label{sec:vmc}Variational Monte Carlo method}
The VMC method is a variational method where we perform the Markov-chain Monte Carlo sampling to optimize the variational parameters such that the variational wave function has the minimum energy. In this section, we explain how to implement the VMC method for our variational wave function.

\subsection{\label{sec:vmc_overview}Overview}
In the variational Monte Carlo method, we estimate an expectation value $\langle {\cal A} \rangle = \frac{\langle \psi |{\cal A}|\psi \rangle}{\langle \psi | \psi \rangle}$ for a given wave function $|\psi \rangle$ by using the Markov-chain Monte Carlo method. To make it clear, we transform $\langle {\cal A} \rangle$ as follows:

\begin{eqnarray}
  \langle {\cal A} \rangle & = & \frac{\langle \psi |{\cal A}|\psi \rangle}{\langle \psi | \psi \rangle} \\
  & = & \int dQ \sum_{\nu} \frac{\langle \psi |Q, \nu \rangle \langle Q,\nu |{\cal A}|\psi \rangle}{\langle \psi | \psi \rangle} \\
  & = & \int dQ \sum_{\nu} \rho(Q,\nu) F \left[ {\cal A}, (Q, \nu) \right].
\end{eqnarray}
Here, we have inserted the completeness relation $\int dQ \sum_{\nu} |Q,\nu \rangle \langle Q,\nu| =1$ in the second line. We choose the real space configuration of electrons $| \nu \rangle = c^{\dagger}_{r_1 \sigma_{1}} c^{\dagger}_{r_2 \sigma_{2}} \cdots c^{\dagger}_{r_N \sigma_{N}} | 0 \rangle$ as the basis on the electron Hilbert space. On the other hand, we choose the normal coordinate configuration $|Q \rangle$ as a basis on the phonon Hilbert space (See Appendix). The probability $\rho(Q,\nu)$ and the quantity $F \left[ {\cal A}, (Q, \nu) \right]$ are defined as
\begin{eqnarray}
  \rho(Q,\nu) = \frac{\left| \langle Q,\nu |\psi \rangle \right|^2}{\langle \psi | \psi \rangle},\\
  F \left[ {\cal A}, (Q, \nu) \right] = \frac{\langle Q, \nu |{\cal A}| \psi \rangle}{\langle Q, \nu | \psi \rangle},
\end{eqnarray}
respectively. For our variational wave function, the inner product is given by
\begin{eqnarray}
  \langle Q, \nu |\psi \rangle = P^{\rm e-ph}(Q,\nu) P^{\rm G}(\nu) P^{\rm J}(\nu)  \langle Q |\psi^{\rm ph}\rangle \langle \nu | \phi^{\rm pair} \rangle.
\end{eqnarray}
By performing the Monte Carlo sampling of $(Q, \nu)$ according to the probability $\rho(Q,\nu)$, we can estimate $\langle A \rangle$ from the average of samples $(Q_i,\nu_i)$:
\begin{eqnarray}
  \langle A \rangle \simeq \frac{1}{N_{\rm MC}}\sum_{i=1}^{N_{\rm MC}} F \left[ {\cal A}, (Q_i, \nu_i) \right],
\end{eqnarray}
where $N_{\rm MC}$ is the number of samples. We usually estimate the energy, its derivatives and some other quantities for the wave function $|\psi_{\alpha} \rangle$ and update the variational parameters by using a minimization method. As the minimization method, we adopt the stochastic reconfiguration (SR) method which enables us to optimize many variational parameters stably\cite{sorella2001,casula2004,sorella2005,sorella2007}.

\subsection{\label{sec:vmc_ip}Inner products}
As shown in the previous subsection, the calculations of the probability $\rho(Q,\nu)$ involves those  of the inner products $\langle \nu | \phi^{\rm pair} \rangle$ and $\langle Q | \psi^{\rm ph}  \rangle$. In this subsection, we present the prescription to calculate these inner products.

For the inner product $\langle \nu | \phi^{\rm pair} \rangle$, it is simply proportional to the determinant of the $(N_{\rm e}/2)\times (N_{\rm e}/2)$ matrix $f_{r_i r_j}$\cite{bouchaud1988,tahara2008}:
\begin{eqnarray}
  \langle \nu | \phi^{\rm pair} \rangle \propto {\rm det}[f_{r_i r_j} ].
\end{eqnarray}
Here, $r_i$ and $r_j$ represent the sites of the $i$-th electron with up spin and the $j$-th electron with down spin, respectively. The calculation of the determinant requires $O(N_{\rm e}^3)$ time.

We next present the expression for $\langle Q | \psi^{\rm ph}  \rangle$. Since the inner product $\langle Q_{\bm q}|m_{\bm q} \rangle$ is nothing but an eigenfunction of a harmonic oscillator, it can be written in terms of the Hermite polynomial $H_n(x)$. Therefore, the inner product $\langle Q | \psi^{\rm ph}  \rangle$ can be written in the following way:
\begin{eqnarray}
  \langle Q |\psi^{\rm ph}\rangle & = & \prod_{\bm q} \left( \sum_{m_{\bm q}=0}^{m_{\bm q}^{\rm max}} c_{m_{\bm q}} \langle Q_{\bm q} | m_{\bm q} \rangle \right) \nonumber \\
  & = & \prod_{\bm q} \left( \sum_{m_{\bm q}=0}^{m_{\bm q}^{\rm max}} c_{m_{\bm q}} N_{m_{\bm q}} H_{m_{\bm q}}({\bar Q}_{\bm q}) \exp(-\frac{{\bar Q}^2_{\bm q}}{2}) \right), \nonumber
\end{eqnarray}
where $N_{n} = ( \sqrt{\pi} 2^n n!)^{-1/2}$ is the normalization factor, which we can calculate in advance and ${\bar Q}_{\bm q}=\sqrt{M \omega_{\bm q}} Q_{\bm q}$ is the dimensionless normal coordinate. In practice, we calculate the values of the Hermite polynomials sequentially according to the following relation:
\begin{eqnarray}
  H_{n}(x) = 2 n H_{n-1}(x) -2(n-1) H_{n-2}(x).
\end{eqnarray}
Thus, the calculation of the inner product takes $O(N m^{\rm max})$ time if $m_{\bm q}^{\rm max}= m^{\rm max}$.

\subsection{\label{sec:vmc_mcupdate}Monte Carlo update schemes}
In this subsection, we describe update schemes of configurations $(Q, \nu)$. We begin with the update of electron configurations. We first choose one of the electrons randomly. In this update, we try to hop the chosen electron from a site $i$ to another site $j$ which is also chosen randomly. If we adopt the standard Metropolis-Hastings algorithm\cite{metropolis1953,hastings1970} for transition probabilities, the acceptance probability $p_{\rm accept}$ is given by
\begin{eqnarray}
  p_{\rm accept} = {\rm min} \left[ \left| \frac{P^{\rm e-ph}(Q, \nu') P^{\rm G}(\nu') P^{\rm J}(\nu') \langle \nu' | \phi^{\rm pair} \rangle}{P^{\rm e-ph}(Q, \nu) P^{\rm G}(\nu) P^{\rm J}(\nu)  \langle \nu | \phi^{\rm pair} \rangle} \right|^2, 1 \right]. \nonumber
\end{eqnarray}
Except for the first update, we do not need to calculate $P^{\rm e-ph}(Q, \nu)$, $P^{\rm G}(\nu)$, $P^{\rm J}(\nu)$, and $\langle \nu | \phi^{\rm pair} \rangle$ again because we have calculated them in the previous update. By updating the values of them, the calculations of $P^{\rm e-ph}(Q, \nu')$, $P^{\rm G}(\nu')$, $P^{\rm J}(\nu')$, and $\langle \nu' | \phi^{\rm pair} \rangle$ can be performed efficiently in the computational time of $O(N)$, $O(1)$, $O(N)$, and $O(N_{\rm e}^2)$\cite{tahara2008}, respectively. We usually repeat this update $N_{e}$-times. Then, we move to the update of the normal coordinates.

Next, we explain updates of the normal coordinates. In this update, we randomly choose one of the normal coordinates $\{Q_{\bm q} \}$ which will be updated. A new candidate $Q_{\bm q}'$ can be generated according to some distribution function $W(Q_{\bm q}')$. If we adopt the Metropolis-Hastings algorithm, the acceptance probability $p_{\rm accept}$ is given by
\begin{eqnarray}
  p_{\rm accept} = {\rm min} \left[ \frac{W(Q_{\bm q})}{W(Q_{\bm q}')} \left| \frac{P^{\rm e-ph}(Q', \nu) \langle Q_{\bm q}'|\psi_{\bm q}^{\rm ph}\rangle}{P^{\rm e-ph}(Q, \nu) \langle Q_{\bm q}|\psi_{\bm q}^{\rm ph}\rangle} \right|^2, 1 \right]. \nonumber
\end{eqnarray}
To calculate $P^{\rm e-ph}(Q', \nu)$, we first transform $\{ Q'\}$ to $\{x'\}$. This can be performed in $O(N)$ time by updating the values of $\{ x \}$. Then, we calculate $P^{\rm e-ph}(Q', \nu)$ in $O(N^2)$ time, while it reduces to $O(N)$, if we keep only short-range part of $\{ \alpha_{ij} \}$. On the other hand, the calculation of $\langle Q_{\bm q}'|\psi_{\bm q}^{\rm ph}\rangle$ requires $O(m_{\bm q}^{\rm max})$ time as described in Sec. \ref{sec:vmc_ip}. As the distribution function $W(Q_{\bm q}')$, we choose the Gaussian distribution $W(Q_{\bm q}') = \frac{1}{\sqrt{2\pi \sigma^2}} \exp(-\frac{Q_{\bm q}'^2}{2\sigma^2})$. Here, the variance $\sigma^2$ is treated as a tuning parameter. We usually repeat this update $N$-times. Then, we perform measurements.

\subsection{\label{sec:vmc_sr}Minimization method}
In this subsection, we briefly review the SR method which is similar to the standard steepest decent (SD) method but enables us to optimize many variational parameters more efficiently and stably.

In both the SD and the SR method, we update variational parameters $\alpha_k$ ($k=1, \cdots, N_{\rm p}$) to 
\begin{eqnarray}
  \alpha'_k = \alpha_k + \delta \alpha_k,
\end{eqnarray}
where 
\begin{eqnarray}
  \delta \alpha_{k} = - \Delta t \sum_{k'=1}^{N_{\rm p}} S^{-1}_{k k'} g_{k'}.  
\end{eqnarray}
Here, $\Delta t$ is a small constant, $S_{kk'}$ is a matrix described below and $g_{k}$ is the energy gradient which is given by
\begin{eqnarray}
  g_k & = & \frac{\partial}{\partial \alpha_k} \frac{\langle \psi_{\alpha} |{\cal H}|\psi_{\alpha} \rangle}{\langle \psi_{\alpha} | \psi_{\alpha} \rangle} \nonumber \\
  & = & 2 \langle {\cal H} {\cal O}_k \rangle - 2 \langle {\cal H} \rangle \langle {\cal O}_k \rangle.
\end{eqnarray}
Here, the operator ${\cal O}_k$ is defined by
\begin{eqnarray}
  {\cal O}_k = \int dQ \sum_{\nu} \left( \frac{1}{\langle Q, \nu | \psi_{\alpha} \rangle} \frac{\partial}{\partial \alpha_k} \langle Q, \nu | \psi_{\alpha} \rangle  \right) | Q, \nu \rangle \langle Q, \nu | \nonumber.
\end{eqnarray}
The difference between the SD and the SR method is the choice of the matrix $S_{k k'}$. In the SD method, we simply choose $S_{k k'}=\delta_{k k'}$. However, a small change in the variational parameters sometimes causes a large change in the variational wave function. This sometimes induces a numerical instability in the optimization. Although we can suppress this instability by taking a sufficiently small $\Delta t$, it slows down the convergence. In the SR method, to suppress this instability, we choose 
\begin{eqnarray}
  S_{k k'} = \langle {\cal O}_k {\cal O}_{k'} \rangle - \langle {\cal O}_k \rangle \langle {\cal O}_{k'} \rangle
\end{eqnarray}
based on the fact that it relates to the squared norm of the variation of the normalized wave function $\Delta^2 = \| | \bar{\psi}_{\alpha}\rangle - | \bar{\psi}_{\alpha+ \delta \alpha} \rangle  \|^2$ as\cite{casula2004,tahara2008}
\begin{eqnarray}
  \Delta^2 = \sum_{k k'} \delta \alpha_k \delta \alpha_{k'} S_{k k'}.
\end{eqnarray}
Here, $| \bar{\psi}_{\alpha}\rangle$ is defined by $| \bar{\psi}_{\alpha}\rangle = | \psi_{\alpha}\rangle/\| | \psi_{\alpha}\rangle \|$. The SR method requires the computational time of $O(N_{\rm p}^3)$ because we need to obtain the inverse matrix $S^{-1}$.

\subsection{\label{sec:vmc_meas}Measured quantities}
As described in Sec. \ref{sec:vmc_sr}, we need to estimate the expectation values $\langle {\cal O}_k \rangle$, $\langle {\cal H} \rangle$, $\langle {\cal H} {\cal O}_k \rangle$, and $\langle {\cal O}_k {\cal O}_l \rangle$ for updating the variational parameters. To estimate them, we measure the quantities  $F \left[ {\cal O}_k,(Q,\nu) \right]$ and $F\left[ {\cal H}, (Q, \nu) \right]$ for each sample $(Q,\nu)$. In this subsection, we present the expression of $F \left[ {\cal O}_k,(Q,\nu) \right]$ and $F\left[ {\cal H}, (Q, \nu) \right]$. Since the expressions for electrons are presented in Ref. \cite{tahara2008}, we focus only on the phonon-related terms.

We first explain the expression of $F\left[ {\cal H}, (Q, \nu) \right]$ which is written as
\begin{eqnarray}
  F\left[ {\cal H}, (Q, \nu) \right] & = & \frac{\langle Q, \nu| {\cal H} | \psi \rangle}{\langle Q, \nu| \psi \rangle} \nonumber \\
  & = & \int dQ' \sum_{\nu'} \langle Q, \nu | {\cal H} |Q', \nu' \rangle \frac{\langle Q',\nu' |\psi \rangle}{\langle Q,\nu |\psi \rangle}. \nonumber
\end{eqnarray}
Among the terms in the Hamilotian ${\cal H}$, the phonon-related terms are the following: the kinetic term $\Pi_{\bm q}^2 = -\frac{\partial^2}{\partial Q_{\bm q}^2}$, the potential term $Q_{\bm q}^2$ and the electron-phonon interaction $g_{ij} x_i n_j$. For the potential term and the electron-phonon interaction term, we obtain  
\begin{eqnarray}
  F \left[ Q_{\bm q}^2, (Q,\nu) \right] & = & Q_{\bm q}^2, \label{eq:pote}\\
  F \left[ g_{ij} x_i n_j, (Q, \nu) \right] & = & g_{ij} x_i n_j. \label{eq:int}
\end{eqnarray}
Since we have already transformed $\{ Q \}$ to $\{ x \}$ during the MC update, we can calculate Eq. (\ref{eq:int}) immediately as well as Eq. (\ref{eq:pote}). For the kinetic term, we obtain
\begin{eqnarray}
  & & F \left[ -\frac{\partial^2}{\partial Q_{\bm q}^2}, (Q,\nu) \right] \nonumber \\
  & = &  \frac{1}{\langle Q, \nu | \psi \rangle} \int dQ' \sum_{\nu'} \langle Q,\nu| -\frac{\partial^2}{\partial Q_{\bm q}^2} |Q',\nu' \rangle \langle Q',\nu' |\psi \rangle \nonumber \\
  & = & - \frac{1}{P^{\rm e-ph}(Q,\nu) \langle Q | \psi^{\rm ph} \rangle} \frac{\partial^2}{\partial Q_{\bm q}^2} P^{\rm e-ph}(Q, \nu) \langle Q| \psi^{\rm ph} \rangle. \nonumber \\
  & = & - \frac{1}{P^{\rm e-ph}(Q,\nu)} \frac{\partial^2 P^{\rm e-ph}(Q, \nu) }{\partial Q_{\bm q}^2} - \frac{1}{\langle Q_{\bm q} | \psi^{\rm ph} \rangle} \frac{\partial^2 \langle Q_{\bm q}| \psi^{\rm ph} \rangle}{\partial Q_{\bm q}^2} \nonumber \\
  & & - \frac{2}{P^{\rm e-ph}(Q,\nu) \langle Q_{\bm q} | \psi^{\rm ph} \rangle} \frac{\partial P^{\rm e-ph}(Q, \nu)}{\partial Q_{\bm q}} \frac{\partial \langle Q_{\bm q} | \psi^{\rm ph} \rangle}{\partial Q_{\bm q}}. \label{eq:kinetic}
\end{eqnarray}
The expressions of $\frac{\partial P^{\rm e-ph}(Q, \nu)}{\partial Q_{\bm q}}$ and $\frac{\partial^2 P^{\rm e-ph}(Q, \nu)}{\partial Q_{\bm q}^2}$ are given by
\begin{eqnarray}
  \frac{\partial P^{\rm e-ph}(Q, \nu)}{\partial Q_{\bm q}} & = & \frac{\partial}{\partial Q_{\bm q}} \exp \left( \frac{1}{\sqrt{N}} \sum_{ij{\bm k}} \alpha_{ij} Q_{\bm k} n_j e^{i{\bm k}\cdot {\bm r}_i}\right) \nonumber \\
  & = & \left( \frac{1}{\sqrt{N}} \sum_{ij} \alpha_{ij} n_j e^{i{\bm q}\cdot {\bm r}_i}  \right) P^{\rm e-ph}(Q, \nu), \nonumber
\end{eqnarray}
and 
\begin{eqnarray}
  \frac{\partial^2 P^{\rm e-ph}(Q, \nu)}{\partial^2 Q_{\bm q}} & = & \left( \frac{1}{\sqrt{N}} \sum_{ij} \alpha_{ij} n_j e^{i{\bm q}\cdot {\bm r}_i}  \right)^2 P^{\rm e-ph}(Q, \nu), \nonumber
\end{eqnarray}
respectively. Since $P^{\rm e-ph}(Q, \nu)$ cancels with the denominator in Eq (\ref{eq:kinetic}), we only have to calculate $\sum_{ij} \alpha_{ij} n_j e^{i{\bm q}\cdot {\bm r}_i}$ which requires $O(N^2)$ time. It reduces to $O(N)$ if we have only short-range ones of $\{ \alpha_{ij} \}$. On the other hand, the expressions of $\frac{\partial \langle Q_{\bm q} | \psi_{\bm q}^{\rm ph} \rangle}{\partial Q_{\bm q}}$ and $\frac{\partial^2 \langle Q_{\bm q} | \psi_{\bm q}^{\rm ph} \rangle}{\partial Q_{\bm q}^2}$ are given by
\begin{eqnarray}
 & & \frac{\partial \langle Q_{\bm q} | \psi_{\bm q}^{\rm ph} \rangle}{\partial Q_{\bm q}} \nonumber \\
 & = & \sum_{m_{\bm q}=0}^{m_{\bm q}^{\rm max}} \lambda_{\bm q} c_{m_{\bm q}} N_{m_{\bm q}} e^{ -{\bar Q}^2_{\bm q}/2 } \left[ \frac{\partial H_{m_{\bm q}}({\bar Q}_{\bm q})}{\partial {\bar Q}_{\bm q}} - {\bar Q}_{\bm q} H_{m_{\bm q}}({\bar Q}_{\bm q}) \right], \nonumber
\end{eqnarray}
and 
\begin{eqnarray}
 & & \frac{\partial^2 \langle Q_{\bm q} | \psi_{\bm q}^{\rm ph} \rangle}{\partial Q_{\bm q}^2} \nonumber \\
 & = & \sum_{m_{\bm q}=0}^{m_{\bm q}^{\rm max}} \lambda_{\bm q}^2 c_{m_{\bm q}} N_{m_{\bm q}} e^{ -{\bar Q}^2_{\bm q}/2 } \nonumber \\
 & & \times \left[ \frac{\partial^2 H_{m_{\bm q}}({\bar Q}_{\bm q})}{\partial {\bar Q}_{\bm q}^2} - 2 {\bar Q}_{\bm q} \frac{\partial H_{m_{\bm q}}({\bar Q}_{\bm q})}{\partial {\bar Q}_{\bm q}} + ({\bar Q}_{\bm q}^2 - 1 ) H_{m_{\bm q}}({\bar Q}_{\bm q}) \right], \nonumber
\end{eqnarray}
respectively. Here, $\lambda_{\bm q}$ is defined by $\lambda_{\bm q} = \sqrt{M \omega_{\bm q}}$. Since we have already calculated $H_{m_{\bm q}}({\bar Q}_{\bm q})$ as well as $\langle Q_{\bm q} | \psi_{\bm q}^{\rm ph} \rangle$ during the MC update, we can calculate its derivatives by using the relations $\frac{\partial H_{n}(x)}{\partial x} = 2n H_{n-1}(x)$ and $\frac{\partial^2 H_{n}(x)}{\partial x^2} = 4n(n -1) H_{n-2}(x)$. The calculations of the derivatives of the inner product require $O(m_{\bm q}^{\rm max})$ time.

Next, we derive the expression for $F \left[ {\cal O}_k,(Q,\nu) \right]$. For the variational parameters $\{ c_{m_{\bm q}} \}$ in the phonon wave function, we obtain
\begin{eqnarray}
  F \left[ {\cal O}_k,(Q,\nu) \right] & = & \frac{1}{\langle Q, \nu | \psi \rangle} \frac{\partial}{\partial c_{m_{\bm q}}} \langle Q,\nu |\psi \rangle \nonumber \\
  & = & \frac{1}{\langle Q|\psi^{\rm ph}\rangle} \frac{\partial}{\partial c_{m_{\bm q}}} \langle Q|\psi^{\rm ph}\rangle \nonumber \\
  & = & \frac{1}{\langle Q|\psi^{\rm ph}\rangle} \frac{\partial}{\partial c_{m_{\bm q}}} \prod_{{\bm q}'} \left( \sum_{m_{ {\bm q}'=0}}^{m_{{\bm q}'}^{\rm max}} c_{m_{{\bm q}'}} \langle Q_{{\bm q}'}| m_{ {\bm q}'} \rangle \right) \nonumber \\
  & = & \frac{\langle Q_{\bm q}|m_{\bm q} \rangle}{\langle Q_{\bm q}|\psi^{\rm ph}_{\bm q}\rangle}. 
\end{eqnarray}
Since the calculations of both the denominator and the numerator have already been performed during the MC updates, we can calculate this quantity immediately. For the variational parameters $\{ \alpha_{ij} \}$ in the electron-phonon correlation factor, we obtain
\begin{eqnarray}
  F \left[ {\cal O}_k,(Q,\nu) \right] & = & \frac{1}{\langle Q, \nu | \psi \rangle} \frac{\partial}{\partial {\alpha}_{ij}} \langle Q,\nu |\psi \rangle \nonumber \\
  & = & \frac{1}{P^{\rm e-ph}(Q, \nu)} \frac{\partial}{\partial \alpha_{ij}} \exp \left( \sum_{i'j'} {\alpha}_{i'j'} x_{i'} n_{j'} \right) \nonumber \\
  & = & x_i n_j,
\end{eqnarray}
which is equal to Eq. (\ref{eq:int}) except for the coefficient $g_{ij}$.

\section{\label{sec:res}Results}
In this section, we present benchmark results to show the efficiency and the accuracy of our variational wave function for electron-phonon coupled systems. In Sec. \ref{sec:res_ed}, we compare our results with that obtained by the ED\cite{marsiglio1995} for the Holstein-Hubbard model\cite{holstein1959} with $4$ site chains. In Sec. \ref{sec:res_gfmc}, to check the validity for larger sizes, we compare our results with that obtained by the Green function Monte Carlo (GFMC)\cite{mckenzie1996} for the Holstein model of spinless fermion. Finally, in Sec. \ref{sec:res_dmrg}, we show that our variational wave function can describe the Peierls CDW state by presenting the data of the charge structure factor.

\subsection{\label{sec:res_ed}Comparison with the exact diagonalization}
In this section, we show comparisons of the ground-state energy between the ED\cite{marsiglio1995} and the VMC for our variational wave function. The Hamiltonian considered here is the Holstein-Hubbard model which is written by
\begin{eqnarray}
  {\cal H} & = & - t \sum_{\langle i,j \rangle, \sigma} (c_{i \sigma}^{\dagger}c_{j \sigma} + {\rm h. c.}) + U \sum_i n_{i \uparrow} n_{i \downarrow}\\ \nonumber
  & & - g \sum_{i} (b^{\dagger}_i + b_i) n_{i} + \omega \sum_i b_i^{\dagger} b_i,
\end{eqnarray}
where $t$, $U$, and $g$ represent the hopping amplitude, the strength of the on-site intraction between electrons and the strength of the electron-phonon interaction, respectively. We consider a one-dimensional system with $N=4$ sites under the periodic boundary condition. In the ED calculations, the phonon Hilbert space was truncated by keeping the basis (Fock states) $| \{ m_{\bm q} \} \rangle $ with the total number of phonons $\sum_{\bm q} m_{\bm q} \leq M^{\rm max}$\cite{marsiglio1995}. The largest value of $M^{\rm max}$ which the author of Ref. \cite{marsiglio1995} took is $M^{\rm max}=40$. In our VMC calculations, we set the cutoff $m_{\bm q}^{\rm max}$ to 40 for all ${\bm q}$ in order to ensure that the possible number of phonons in the VMC is not smaller than that in the ED. We confirmed that this cutoff is sufficiently large to represent the physical limit $m_{\bm q}^{\rm max} \rightarrow \infty$.

\begin{figure}[h*]
  \includegraphics[width=9cm]{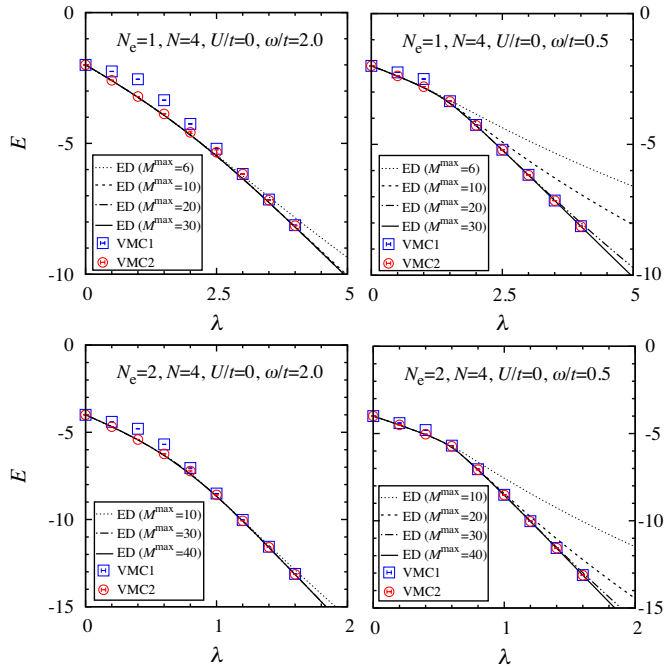}
\caption{(Color online) Comparisons of ground-state energies between results obtained from ED \cite{marsiglio1995} (lines) and VMC (colored open symbols) at $U=0$. VMC1 and VMC2 represent the VMC results without and with ${\cal P}^{\rm e-ph}$, respectively.}
\label{fig:hlstModel}
\end{figure}

Figure \ref{fig:hlstModel} shows comparisons of the ground-state energies $E$ between the ED and the VMC at $U=0$. We consider the cases of a single electron ($N_{\rm e}=1$) or two electrons with opposite spins ($N_{\rm e}=2$) at $\omega/t=2.0$ and 0.5. In the horizontal axis, a dimensionless quantity $\lambda$ is defined by $\lambda = g^2/(2 t \omega)$. In these figures, we have shown two kinds of VMC results obtained by the variational wave function with the electron-phonon correlation factor ${\cal P}^{\rm e-ph}$ and without it, respectively. Even without ${\cal P}^{\rm e-ph}$, our variational wave function can describe the ground-state wave function precisely at $\lambda=U=0$ or at $t=0$ where electrons are localized. We see good agreement between the results of the VMC without ${\cal P}^{\rm e-ph}$ and the ED for large $\lambda$ as well. This is because electrons are self-trapped and nearly localized\cite{hohenadler2004}. For moderate values of $\lambda$, however we clearly observe discrepancies between the VMC without ${\cal P}^{\rm e-ph}$ and the ED. In this intermediate coupling region, we find that the introduction of ${\cal P}^{\rm e-ph}$ remarkably improves the accuracy of the ground-state energy as seen in the figures.

\begin{figure}[H]
  \includegraphics[width=8cm]{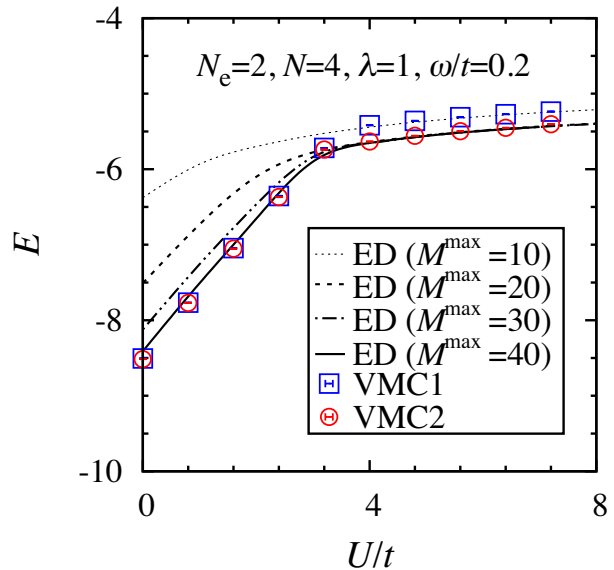}
\caption{(Color online) Comparisons of $U$ dependence of ground-state energies between ED\cite{marsiglio1995} (lines) and VMC (colored open symbols) for two electrons at ($\lambda, \omega/t$)=(1, 0.2). VMC1 and VMC2 represent the VMC results without and with ${\cal P}^{\rm e-ph}$, respectively.}
\label{fig:hlstHbModel}
\end{figure}

In Fig. \ref{fig:hlstHbModel}, we compare the results of the VMC with the ED for $U \geq 0$. We consider two electrons at $\lambda=1.0$ and $\omega=0.2t$. For small values of $U < 4.0$, the ground-state energies obtained by the ED have not converged yet as a function of the cutoff $M^{\rm max}$. Since the definition of the cutoff in the number of phonons in the ED is different from that in the VMC, it is allowed that the ground-state energies by the VMC are lower than those by the ED. For $U\geq 4.0$, the ground-state energies obtained by the ED have converged well and those by the VMC well reproduce them if we include ${\cal P}^{\rm e-ph}$ in the variational wave function.

We next show how the results depend on the range of the variational parameters $\alpha_{ij}$. In Table \ref{tab:cutoff_dep}, we present an example of cutoff dependence of the ground-state energy. By introducing the electron-phonon correlation factor ${\cal P}^{\rm e-ph}$ only with the $i=j$ part of $\alpha_{ij}$, we have already obtained the result with a few percent accuracy (``$r_{\rm c}$=0'' in Table \ref{tab:cutoff_dep}). The accuracy of the ground-state energy can be slightly improved if we include the nearest-neighbor-sites part of $\alpha_{ij}$ (``$r_{\rm c}$=1'' in Table \ref{tab:cutoff_dep}). However, we did not find improvement within the error bar even if we increase the cutoff in the range of $\alpha$(``$r_{\rm c}$=2'' in Table \ref{tab:cutoff_dep}).

\begin{table}[H]
\begin{center}
\scalebox{1.15}{
\footnotesize
\begin{tabular}{lcccccc}
\hline \hline
      & No $r_{\rm c}$ & $r_{\rm c}$=0 & $r_{\rm c}$=1 & $r_{\rm c}$=2 & ED \rule{0pt}{3.0ex} \\ \hline
 Energy & -2.501(1) & -2.775(2) & -2.795(1) & -2.796(1) & -2.8343 \rule{0pt}{3.0ex}  \rule[-1.2ex]{0pt}{0pt}  \\
\hline \hline
\end{tabular}
}
\caption{Range $r_{\rm c}$ dependence of VMC ground-state energy along with ED ground-state energy\cite{marsiglio1995} at $(N_{\rm e}, N, U/t, \omega/t, \lambda)$ = (1, 4, 0, 0.5, 1).  Here, the range $r_c$ (in the unit of the lattice constant) represents the cutoff in the varitaional parameter $\alpha_{ij}$. We keep $\alpha_{ij}$ only within the distance $r_{ij} \leq r_{\rm c}$ in the electron-phonon correlation factor ${\cal P}^{\rm e-ph}$. ``No $r_{\rm c}$'' indicates the result obtained by the VMC without ${\cal P}^{\rm e-ph}$.}
\label{tab:cutoff_dep}
\end{center}
\end{table}

Finally, we discuss the structure of our variational wave function to clarify the role of the electron-phonon correlation factor. For simplicity, we consider only the $i=j$ part of $\alpha_{ij}$ in the electron-phonon correlation factor and the case of $N_{\rm e}=2$. In addition, we focus on the zero-phonon state $|0\rangle_{\rm ph}$ in the phonon variational wave function which is most relevant in the antiadiabatic regime. In this situation, the focused part of our variational wave function is written as
\begin{eqnarray}
  e^{\sum_{l} \alpha (b_l^{\dagger} + b_l) n_l} \left[ | 0 \rangle_{\rm ph} \left(  \sum_{i,j=1}^{N} f_{ij} c_{i \uparrow}^{\dagger} c_{j \downarrow}^{\dagger} \right) |0 \rangle_{\rm ele}  \right] \nonumber \\
  = \sum_{i,j=1}^{N} f_{ij} \left[ e^{\alpha (b_i^{\dagger} + b_i)} c_{i \uparrow}^{\dagger}\right] \left[ e^{\alpha (b_j^{\dagger} + b_j)} c_{j \downarrow}^{\dagger} \right] | 0 \rangle. 
\end{eqnarray}
Here, $\alpha$ is the variational parameter, $b_{l}^{\dagger}$ indicates the creation operator of a phonon at the site $l$, $|0 \rangle_{\rm ele}$ is the electron vaccuum state, and $|0 \rangle$ is defined by $|0 \rangle = |0 \rangle_{\rm ph} |0 \rangle_{\rm ele}$. In the above equation, we can replace the operator $e^{\alpha (b_i^{\dagger} + b_i)}$ by the displacement operator $e^{\alpha (b_i^{\dagger} - b_i)}$ with an additional factor $e^{\alpha^2}$. This is immediately seen from the relations $e^{\alpha (b_i^{\dagger} + b_i)} = e^{\alpha^2/2} e^{\alpha b_i^{\dagger}} e^{\alpha b_i}$, $e^{\alpha (b_i^{\dagger} - b_i)} = e^{-\alpha^2/2} e^{\alpha b_i^{\dagger}} e^{-\alpha b_i}$, and $e^{\alpha b_i}|0 \rangle =|0\rangle$. The first and the second relations are obtained from the Baker-Campbell-Hausdorff formula. The additional factor $e^{\alpha^2}$ is unimportant, because we can remove it by redefining the variational parameter $f_{ij}$ by ${\tilde f}_{ij} = f_{ij} e^{\alpha^2}$. Thus, it turns out that the focused part of our variational wave function is written as
\begin{eqnarray}
   \sum_{i,j=1}^{N} {\tilde f}_{ij} {\tilde c}_{i \uparrow}^{\dagger} {\tilde c}_{j \downarrow}^{\dagger}  | 0 \rangle. 
\end{eqnarray}
Here, ${\tilde c}_{i \sigma}^{\dagger} = e^{\alpha (b_i^{\dagger} - b_i)} c_{i \sigma}^{\dagger}$ describes an electron tied to the lattice displacement, i.e. a polaron. This transformation is known as the Lang-Firsov transformation\cite{lang1962}. If we include the short-range part of $\alpha_{ij}$, the varitational wave function clearly takes into account the effect of an electron displacing the neighbor lattices in addition to the lattice where it occupies. For systems with strong local (Holstein-type) electron-phonon interactions, it has been numerically shown that the electron-displacement correlation function decays exponentially at long distance\cite{jeckelmann1998,wellein1998,bonca1999,ku2002}. In such a case, long-range part of $\alpha_{ij}$ should be negligible.

\subsection{\label{sec:res_gfmc}Size dependence}
In the previous section, we have shown benchmarks for small systems with only 4 sites. In order to check the accuracy of our variational wave function for larger systems, we compare the VMC results with that obtained by the GFMC\cite{mckenzie1996}. The model considered here is the Holstein model of spinless fermions defined by

\begin{eqnarray}
  {\cal H} & = & -t \sum_{\langle i,j \rangle} (c^{\dagger}_i c_j + {\rm h. c.}) \nonumber \\
  & & - g \sum_{i} (b^{\dagger}_i + b_i) (n_i-\frac{1}{2}) + \omega \sum_i b_i^{\dagger} b_i. \label{eq:ham2}
\end{eqnarray}
We consider one-dimensional systems with $N$ sites at half filling. The periodic/anti-periodic boundary condition is applied, if the number of fermions is odd/even. For the Tomonaga-Luttinger-liquid (TLL) in the conformally invariant system under these boundary conditions, the ground-state energy $E(N)$ scales in the leading order as\cite{blote1986}
\begin{eqnarray}
  \frac{E(N)}{N} = \epsilon_{\infty} - \frac{\pi u_{\rho}}{6 N^2},
\end{eqnarray}
where $\epsilon_{\infty}$ is the ground-state energy density of the infinite systems and $u_{\rho}$ is the velocity of charge excitations.

\begin{figure}[H]
\begin{center}
  \includegraphics[width=8cm]{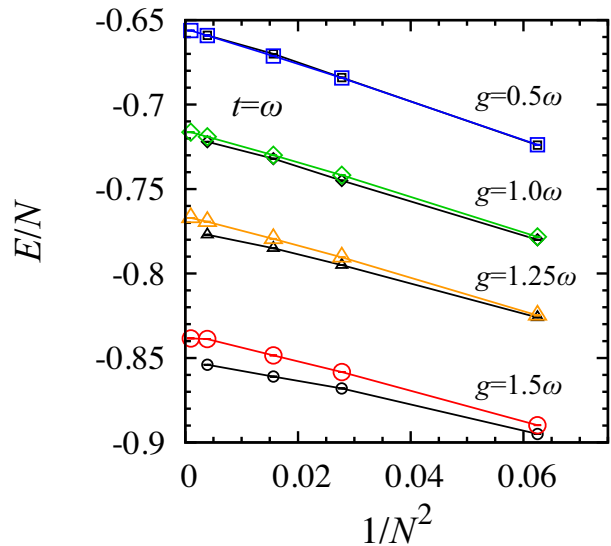}
\caption{(Color online) Comparisons of the ground-state energy density between the VMC with ${\cal P}^{\rm e-ph}$ (colored large symbols) and GFMC\cite{mckenzie1996} (black small symbols). We set the cutoff $m_{\bm q}^{\rm max}$ to 10 for all ${\bm q}$ and confirmed that the quantities have already well converged as a function of $m_{\bm q}^{\rm max}$.}
\label{fig:energy_GFMC}
\end{center}
\end{figure}

In Fig. \ref{fig:energy_GFMC}, we show the comparisons of the ground-state energy density in the TLL phase. According to the DMRG study, the Peierls transition occurs at $g/\omega = 1.61(1)$\cite{bursill1998}. For $N=4$, we observe the good agreements between the VMC and the GFMC for all values of $g$ presented here. For larger systems with $N=6, 8, 16$, we clearly see that the deviation of the VMC results from the GFMC results becomes larger as the value of $g$ increases. However, the discrepancy is still within 2 $\%$.

In order to show how the correlation factor ${\cal P}^{\rm e-ph}$ improves the accuracy of the ground-state energy, we present the data obtained by the VMC without ${\cal P}^{\rm e-ph}$ (VMC1) and the VMC with it (VMC2) along with that by the GFMC\cite{mckenzie1996} in Table \ref{tab:gfmc}. From these data, we see that the introduction of ${\cal P}^{\rm e-ph}$ improves the accuracy of the ground-state energy significantly for all system sizes presented here.

\begin{table}[H]
\begin{center}
\scalebox{1.25}{
\footnotesize
\begin{tabular}{lccccc}
\hline \hline
      & $N$=4 & $N$=6 & $N$=8 & $N$=16  \rule{0pt}{2.6ex} \\ \hline
 VMC1 &  -0.7832(7) & -0.7749(2) & -0.7739(2) & -0.7732(4) \rule{0pt}{2.6ex} \\
 VMC2 & -0.8904(5) & -0.8583(3) & -0.8484(3) & -0.8388(5) \rule{0pt}{2.6ex} \\
 GFMC & -0.895(1) & -0.868(1) & -0.861(2) & -0.854(1) \rule{0pt}{2.6ex} \rule[-1.2ex]{0pt}{0pt} \\
\hline \hline
\end{tabular}
}
\caption{Ground-state energy densities obtained by VMC without ${\cal P}^{\rm e-ph}$(VMC1), VMC with it (VMC2), and GFMC\cite{mckenzie1996}. The system parameters are chosen at $t=\omega$ and $g=1.5 \omega$.}
\label{tab:gfmc}
\end{center}
\end{table}

\subsection{\label{sec:res_dmrg}Charge structure factor in the Peierls CDW state}
The Peierls CDW state is a state which is realized owing to the electron-phonon coupling. In order to show that our variational wave function can describe the Peierls CDW state, we present results of the charge structure factor in this subsection. We check its accuracy by comparing with that obtained by the DMRG\cite{fehske2005}. For this comparison, we consider the spinless Holstein model [Eq.(\ref{eq:ham2})] at half filling again. The applied boundary condition is also the same as that in the previous subsection. The measured charge structure factor is defined by
\begin{eqnarray}
  S_{c}(\pi)  = \frac{1}{N^2} \sum_{i,j} (-1)^j \langle (n_i -\frac{1}{2}) (n_{i+j} -\frac{1}{2}) \rangle .
\end{eqnarray}

\begin{figure}[H]
\begin{center}
  \includegraphics[width=9cm]{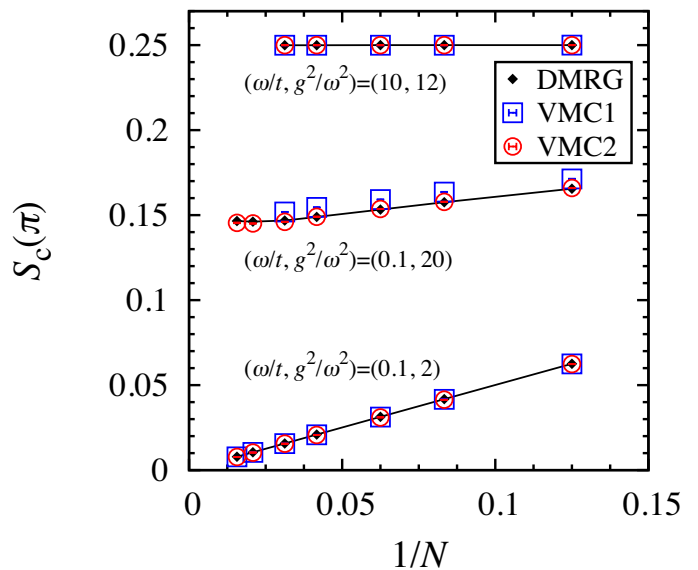}
\caption{(Color online) Comparisons of the charge structure factor $S_{c}(\pi)$ between the VMC (colored open symbols) and DMRG\cite{fehske2005} (black diamonds). The VMC1 and VMC2 represent the VMC results without and with ${\cal P}^{\rm e-ph}$, respectively. For $(\omega/t, g^2/\omega^2)=(0.1,2)$, we set the cutoff $m_{\bm q}^{\rm max}$ to 20 for all $q$. For $(\omega/t, g^2/\omega^2)=(0.1,20)$/(10,12), we set $m_{\bm q}^{\rm max}$ to 360/80 for ${\bm q}={\bm \pi}$ and to 10 for others.}
\label{fig:stf_dmrg}
\end{center}
\end{figure}

In Fig. \ref{fig:stf_dmrg}, we show the comparison of $S_{c}(\pi)$ between the DMRG\cite{fehske2005} and the VMC for three different parameter sets. The parameter set $(\omega/t, g^2/\omega^2)=(0.1,2)$ belongs to the TLL phase, and the other parameter sets $(\omega/t, g^2/\omega^2)=(10,12)$ and (0.1,20) belong to the Peierls CDW phase\cite{bursill1998}. In the TLL phase, the charge structure factor decreases as the system size increases, and eventually vanishes in the thermodynamic limit. Both the DMRG and VMC results are consistent with this behavior. The agreement between the DMRG and VMC results (with or without ${\cal P}^{\rm e-ph}$) is excellent as seen in the figure.

In the CDW phase at $(\omega/t, g^2/\omega^2)=(10,12)$, the VMC results agree with the DMRG results even without ${\cal P}^{\rm e-ph}$. In this case, the charge structure factor does not show clear system-size dependence and takes the value close to the maximum value 1/4, indicating a strong charge order. In contrast, the charge order is weaker for $(\omega/t, g^2/\omega^2)=(0.1,20)$. In this intermediate regime, the VMC result without ${\cal P}^{\rm e-ph}$ clearly deviates from the DMRG result. However, the accuracy of the VMC results improves by including ${\cal P}^{\rm e-ph}$.

\section{\label{sec:sum}SUMMARY AND OUTLOOK}
We have developed a VMC method for electron-phonon coupled systems with the diagonal electron-phonon interactions. The proposed variational wave function includes a correlation factor which takes into account the effect of the entanglement between electrons and phonons. By comparing the VMC results with the previous results obtained by the ED, the GFMC and the DMRG, we have shown that the correlation factor significantly improves the accuracy of the ground-state energy as well as the correlation function (the charge structure factor). Compared with the other numerical methods, the advantage of the VMC method is its wide applicability and flexibility. We do neither have restrictions on the forms of (diagonal) electron-phonon interactions nor lattice structures (spatial dimensionality). Furthermore, our variational wave function can describe dispersive phonons as well as dispersionless (Einstein) phonons. It is also possible to treat several different branches of phonon dispersions. Therefore, we expect that our approach will open a way to study a wide variety of electron-phonon coupled systems. For example, the present method will enable us to tackle the challenging issues such as phonon effects on superconductivity in strongly correlated electron systems and the competition between the charge/spin orders and superconductivity. So far, we have considered systems with the diagonal electron-phonon interactions. The extension to systems with off-diagonal electron-phonon interactions as well as diagonal ones is an intriguing challenge.

\section*{ACKNOWLEDGMENTS}%*'ð''¯'é'ƁAæ"ª'ɔԍ†'ª''©'È'¢B
The code was developed based on the VMC code for electron systems (without electron-phonon couplings) which was implemented by D. Tahara and S. Morita. We thank K. Ido for continuous discussion. We appreciate H. Fehske and A. R. Bishop for providing their DMRG data, F. Marsiglio for providing his ED data and R. H. McKenzie for allowing us to reproduce their GFMC data. This work is financially supported by MEXT HPCI Strategic Programs for Innovative Research (SPIRE)(grant number 130007) and Computational Materials Science Initiative (CMSI). This work was also supported by Grant-in-Aid for
Scientific Research {(No. 22104010, and No. 22340090)} from MEXT, Japan.

%\bibliography{reference}% Produces the bibliography via BibTeX.

\appendix

\section{Precise treatment of  the normal coordinates}
In this paper, we have assumed that the normal coordinate operators $\{ Q_{\bm q} \}$ are Hermitian to simplify the explanations. However, the normal coordinate operators are non-Hermitian in the presence of the electron-phonon coupling. In this case, the eigenstates of the normal coordinates do not form a complete basis of the phonon Hilbert space. Therefore, in this appendix, we formulate the precise treatment of the normal coordinates.

The normal coordinate $Q_{\bm q}$ and its conjugate momentum $\Pi_q$ are defined by
\begin{eqnarray}
  Q_{\bm q} & = & \frac{1}{\sqrt{N}} \sum_{j} x_j e^{-i {\bm q} \cdot {\bm r}_j}, \\
  \Pi_{\bm q} & = & \frac{1}{\sqrt{N}} \sum_{j} p_j e^{i {\bm q} \cdot {\bm r}_j},
\end{eqnarray}
respectively. These operators are non-Hermitian except for the wave vectors ${\bm q}^{\ast}$ satisfiying $\sin({\bm q}^{\ast} \cdot {\bm r}_j)=0$ for all $j$ due to the imaginary part of the exponential. In one dimension, $q^{\ast}$ = $0, \pi$. In order to avoid treating these non-Hermitian operators, we instead treat the following Hermitian operators: 

\begin{eqnarray}
  Q_{\bm q}^{\rm R} & = & \sqrt{\frac{2}{N}} \sum_{j} x_j \cos(- {\bm q} \cdot {\bm r}_j), \\
  Q_{\bm q}^{\rm I} & = & \sqrt{\frac{2}{N}} \sum_{j} x_j \sin(- {\bm q} \cdot {\bm r}_j), \\
  \Pi_{\bm q}^{\rm R} & = & \sqrt{\frac{2}{N}} \sum_{j} p_j \cos(- {\bm q} \cdot {\bm r}_j),\\
  \Pi_{\bm q}^{\rm I} & = & \sqrt{\frac{2}{N}} \sum_{j} p_j \sin(- {\bm q} \cdot {\bm r}_j).
\end{eqnarray}
Since we have the relations $Q_{\bm q}^{\rm R}=Q_{-{\bm q}}^{\rm R}$, $Q_{\bm q}^{\rm I}=-Q_{-{\bm q}}^{\rm I}$, $\Pi_{\bm q}^{\rm R}=\Pi_{-{\bm q}}^{\rm R}$, $\Pi_{\bm q}^{\rm I}=-\Pi_{-{\bm q}}^{\rm I}$, we restrict them to those with the wave vector to half of the first Brillouin zone. We can easily check that these operators satisfy the canonical commutation relations:
\begin{eqnarray}
 \left[ Q_{\bm q}^{\rm R}, \Pi_{{\bm q}'}^{\rm R} \right]   & = &  i\delta_{{\bm q},{\bm q}'}, \\
 \left[ Q_{\bm q}^{\rm I}, \Pi_{{\bm q}'}^{\rm I} \right]   & = &  i \delta_{{\bm q},{\bm q}'}.
\end{eqnarray}
In terms of these Hermitian operators, we can rewrite ${\cal H}^{\rm ph}$ as
\begin{eqnarray}
  {\cal H}^{\rm ph} & = & \frac{1}{2M} \sum_{{\bm q}^{\ast}  } \left( \Pi_{{\bm q}^{\ast}}^2 + M^2 \omega_{{\bm q}^{\ast}}^2 Q_{{\bm q}^{\ast}}^2 \right) \nonumber \\
  & & + \frac{1}{2M} \sum_{\bm q} {}^{'} \left[ (\Pi_{\bm q}^{\rm R})^2 + M^2 \omega_{\bm q}^2 (Q_{\bm q}^{\rm R})^2 \right] \nonumber \\
  & & + \frac{1}{2M} \sum_{\bm q} {}^{'} \left[ (\Pi_{\bm q}^{\rm I})^2 + M^2 \omega_{\bm q}^2 (Q_{\bm q}^{\rm I})^2 \right],
\end{eqnarray}
where the prime on the summation means the summation over half of the first Brillouin zone except for ${\bm q}={\bm q}^{\ast}$. Correspondingly, we modify the phonon variational wave function as follows:
\begin{eqnarray}
  | \psi^{\rm ph} \rangle & = & \prod_{ {\bm q}^{\ast} } \left( \sum_{m_{{\bm q}^{\ast}}=0}^{m_{{\bm q}^{\ast}}^{\rm max}} c_{m_{{\bm q}^{\ast}}} | m_{{\bm q}^{\ast}} \rangle \right) \prod_{\bm q} {}^{'} \left( \sum_{m_{\bm q}^{\rm R}=0}^{m_{\bm q}^{{\rm R}, \rm max}} c_{m_{\bm q}^{\rm R}} | m_{\bm q}^{\rm R} \rangle \right) \nonumber \\
  & & \otimes \prod_{\bm q} {}^{'} \left( \sum_{m_{\bm q}^{\rm I}=0}^{m_{\bm q}^{{\rm I}, \rm max}} c_{m_{\bm q}^{\rm I}} | m_{\bm q}^{\rm I} \rangle \right).
\end{eqnarray}
Here, $| m_{\bm q}^{\rm R(I)} \rangle$ represents a Fock state which is related to the normal coordinate $Q_{\bm q}^{\rm R(I)}$. The prime on the product means the the product over half of the first Brillouin zone except for $\{ {\bm q}^{\ast}\}$. We adopt the eigenstates of these (modified) normal coordinate operators $\{ Q_{{\bm q}^{\ast}} \}$, $\{ Q_{\bm q}^{\rm R} \}$, $\{ Q_{\bm q}^{\rm I} \}$ as a complete basis of the phonon Hilbert space. Even after these modifications, we can still treat these normal coordinate operators on an equal footing. Therefore, the procedure in the VMC method does not change.

\end{document}